\documentclass[prd,twocolumn,amsmath,amssymb,floatfix]{revtex4}
\usepackage{bm}
\usepackage{amsmath}
\usepackage{amssymb}
\usepackage{epsf}
\usepackage{color}
\usepackage{natbib}
\usepackage{graphicx}

\newcommand{\OmegaMnow}{\Omega_{{\rm M,0}}}
\newcommand{\gnewton}{{$g_{\rm Newton}$}}
\newcommand{\gfr}{{$g_{f(R)}$}}

\newcommand{\ackntext}[1]{#1}

\newcommand{\lsc}{\mathcal{L}}

\begin{document}
\title{Non-linear evolution of $f(R)$ cosmologies I: methodology}
\author{Hiroaki Oyaizu}
\email{oyachai@uchicago.edu}
\affiliation{
Department of Astronomy \& Astrophysics, University of Chicago, Chicago IL 60637\\
Kavli Institute for Cosmological Physics, University of Chicago, Chicago IL 60637\\
}
\date{\today}

\begin{abstract}
We introduce the method and the implementation of a cosmological simulation of 
a class of metric-variation $f(R)$ models that accelerate the cosmological expansion
without a cosmological constant and evade solar-system bounds of small-field
deviations to general relativity.
Such simulations are shown to reduce to solving a non-linear Poisson equation
for the scalar degree of freedom introduced by the $f(R)$ modifications.
We detail the method to efficiently solve the non-linear Poisson equation by using 
a Newton-Gauss-Seidel relaxation scheme coupled with multigrid method to 
accelerate the convergence.
The simulations are shown to satisfy tests comparing the simulated outcome to
analytical solutions for simple situations, and the dynamics of the simulations
are tested with orbital and Zeldovich collapse tests.
Finally, we present several static and dynamical simulations using realistic 
cosmological parameters to highlight the differences between standard physics
and $f(R)$ physics.
In general, we find that the $f(R)$ modifications result in stronger
gravitational attraction that enhances the dark matter power spectrum
by $\sim 20\%$ for large but observationally allowed $f(R)$ modifications.
More detailed study of the non-linear $f(R)$ effects on the power spectrum
are presented in a companion paper.
\end{abstract}

\maketitle

%%%%%%%%%%%%%%%%%%%%%%%%%%%%%%%%%%%%%%%%%
\section{Introduction} \label{sec:intro}
%%%%%%%%%%%%%%%%%%%%%%%%%%%%%%%%%%%%%%%%%

Explaining the accelerated cosmological expansion has become one of the
greatest challenges in modern day physics \citep[see e.g.,][]{frieman08a}.
Of the many potential explanations, two classes are popular today: ones 
that propose new and unseen
form of energy with negative pressure, and ones that modify the Einstein-Hilbert
action.
Of particular interest in recent literature are a class of modifications to the Einstein
action by an addition of non-linear functions of the Ricci scalar, $R$, which have been
demonstrated to cause accelerated expansion for a range of $f(R)$ functions
\citep{carroll04a,nojiri03a,capozziello03a}.

In addition to explaining the cosmic acceleration, such $f(R)$ modifications must satisfy stringent
constraints imposed by solar system tests \citep{williams96a,anderson98a,bertotti03a,williams04a} 
as well as cosmological constraints
imposed by recent dark energy measurements \citep{knop03a,riess04a,spergel07a}.
Recently, a new class of $f(R)$ models was proposed by Hu \& Sawicki \citep[][hereafter HS model]{hu07a}
 that has the flexibility to reproduce the observed cosmological acceleration as 
 well as to satisfy solar system bounds on deviations from the Newtonian limits
 of general relativity.
In the proposed model, the scalar degree of freedom introduced by the $f(R)$ 
modification becomes massive in regions of high curvature and is hidden 
by the so-called chameleon mechanism \citep{khoury04a,khoury04b}.
 
 In such models that evade both the cosmological background test and solar
 system tests, the next strongest constraint comes from the effects of the $f(R)$
 modifications to the intermediate lengths scales, in the regime of cosmological
 large scale structures.
 Outside of dense dark matter halos, the light scalar degree of freedom produces
 a long range fifth force that enhances gravitational attraction and thereby potentially
 change clustering of matter.
 Such effects were studied in the linear regime \citep{song07a,hu07a,pogosian08a},
 in which large deviations of dark matter power spectrum were obtained even 
 for small $f(R)$ modifications.
 Furthermore, an attempt to model some of the non-linear chameleon physics using
 a parametrized framework on top of a halo model are presented in \cite{hu07b}.
 However, those studies remain inconclusive because the effects of non-linear growth
 of structure and the chameleon mechanism (which requires non-linear growth) are
 not included in the analyses.
 For such non-linear study, a realistic N-body simulation of structure formation in $f(R)$
 cosmologies is needed.
 
In this paper, we introduce a formalism and an implementation of a cosmological N-body
simulation with fully non-linear treatment of $f(R)$ modifications.
Although somewhat limited in the attainable dynamic range, we show
that our implementation can capture the essential non-linear features of the HS model
that have large impact on the cosmological large scale structure formation.
In a companion paper, we apply our methodology to realistic cosmological 
simulations and discuss the implications of the $f(R)$ model on the large scale 
structure of the Universe.
 
This paper is organized as follows.  In \S \ref{sec:fr}, we review
the general properties of $f(R)$ cosmology and give details on 
the particular model of $f(R)$ that we choose.
The numerical algorithms to solve for the $f(R)$ cosmological 
large scale structure formation are developed in \S \ref{sec:numerical},
and are tested in \S \ref{sec:tests}.
Several test cases under the HS model, highlighting the different
aspects of the $f(R)$ model, are presented in \S \ref{sec:results}.
In addition, a brief discussion of full cosmological simulations with $f(R)$
modifications are presented in \S \ref{sec:results}; a more complete
discussion will be presented in a companion paper.
Finally, conclusions and discussions of possible future extension of
our formalism, are given in \S \ref{sec:conclusion}.

%%%%%%%%%%%%%%%%%%%%%%%%%%%%%%%%%%%%%%%%%
\section{$f(R)$ cosmology} \label{sec:fr}
%%%%%%%%%%%%%%%%%%%%%%%%%%%%%%%%%%%%%%%%%
The so-called $f(R)$ theories are modifications to general relativity (GR) given by the 
following action:
\begin{eqnarray}
S =  \int{d^4 x \sqrt{-g} \left\{ \frac{R+f(R)}{2 \kappa^2} + L_{m} \right\} }
\end{eqnarray}
where $R$ is the scalar curvature (i.e., the Ricci scalar), $f(R)$ is an 
arbitrary function of $R$, $\kappa^2 = 8 \pi G$, and $L_{m}$ is the 
Lagrangian of the ordinary matter.
Setting $f(R)=0$ recovers ordinary GR without a cosmological constant.
We follow the standard procedure of setting $c=1$ and $\hbar=1$ throughout
the paper unless otherwise noted.
The variation of the action with respect to the metric results in the field equation
for $g_{\alpha \beta}$, 
\begin{eqnarray}
G_{\alpha \beta} + f_R R_{\alpha \beta} - \left(\frac{f}{2} - \Box f_R \right)g_{\alpha \beta} \nonumber \\
- \nabla_{\alpha} \nabla_{\beta} f_R = \kappa^2 T_{\alpha \beta} \label{eqn:gfield}
\end{eqnarray}
where $f_{R} \equiv \frac{d f(R)}{dR}$, $G_{\alpha \beta}$ is the unmodified
Einstein tensor, and $T_{\alpha \beta}$ is the energy-momentum
tensor.
We work with the standard assumption that the Universe is filled with 
a perfect fluid of the form 
\begin{eqnarray}
T_{\alpha \beta} = (\rho + P)U_{\alpha} U_{\beta} + P g_{\alpha \beta},
\end{eqnarray}
where $\rho$ and $P$ are the fluid energy density and pressure
and $U$ is the four velocity of the fluid.
Because we are only concerned with modeling the effects of late time
accelerated expansion, we neglect the relativistic energy component
and thus set $P = 0$.
For the background Friedmann-Robertson-Walker metric, the modified
Einstein equation becomes the modified Friedmann equation
\begin{eqnarray}
H^2 - \bar{f}_{R}(H H' + H^2) + \frac{\bar{f}}{6} + H^2 \bar{f}_{RR} \bar{R}' = \frac{\kappa \bar{\rho}}{3},
\end{eqnarray}
where $H$ is the Hubble parameter, $\bar{f}_{RR}$ is the second derivative of
$\bar{f}$ with respect to $R$, and $'$ denote derivatives with respect to $\ln a$.
The over-bars signify that the quantities are defined at the cosmological background
level.
The trace of Eqn. (\ref{eqn:gfield}) becomes a dynamical equation for $f_{R}$:
\begin{eqnarray}
3 \Box f_{R} - R + f_{R}R - 2f = -\kappa^2 \rho. \label{eqn:field}
\end{eqnarray}

Recently, a phenomenologically viable model of $f(R)$ was proposed in 
\cite{hu07a,hu07b}, in which  
\begin{eqnarray}
f(R) = -m^2 \frac{c_{1}(R/m^2)^n}{c_{2}(R/m^2)^n + 1} ,
\end{eqnarray}
where $c_1$, $c_2$ and $n$ are free dimensionless
parameters and 
\begin{eqnarray}
m^2 \equiv \frac{8 \pi G \bar{\rho}_{\rm M}}{3}.
\end{eqnarray}
This form of $f(R)$ can be tuned to match the $\rm \Lambda$CDM expansion history 
while not introducing a true cosmological constant.
In addition, the above form has the potential to evade solar system tests via
the so-called chameleon mechanism \citep{khoury04a,khoury04b},
thus remaining viable even with the stringent constraints placed by such
tests.
However, the HS model has interesting deviations from $\Lambda$CDM in the 
intermediate scales that result in distinct signatures in large scale structure
formation.
Such deviations have been studied in the linear regime \citep{song07a,song07b,pogosian08a},
but the lack of insights into the modification's behavior in the non-linear regime
places a strong limitation on the applicability of comparison of such studies to observations.

The free parameters, $c_1$, $c_2$, and $n$, are chosen so that the resulting background
expansion history closely matches to that of $\Lambda$CDM.  
This requirement is equivalent to requiring the background $f_R$ field strength today, $\bar{f}_{R0}$,
to be sufficiently small ($\bar{f}_{R0} \ll 1$) or, alternatively, $R_{0} \gg m$.
In this parameter regime, the field is always near the minimum of the effective
potential,
\begin{eqnarray}
\bar{R} &=& 8 \pi G \bar{\rho}_{\rm M} - 2 \bar{f} \approx 8 \pi G \bar{\rho}_{\rm M} + 2\frac{c_1}{c_2}m^2.
\end{eqnarray}
In the limit of $\Lambda$CDM, we have
\begin{eqnarray}
\bar{R}_{\rm \Lambda CDM} &=& 3 m^2 \left( a^{-3} + 4 \frac{\Omega_{\Lambda}}{\Omega_{\rm M}} \right),
\end{eqnarray}
and thus, to approximate the $\Lambda$CDM expansion history, we set 
\begin{eqnarray}
\frac{c_1}{c_2} = 6 \frac{\Omega_{\Lambda}}{\Omega_{\rm M}}.
\end{eqnarray}
Two free parameters remain, namely $n$ and $c_1/c_2^2$, that control how 
closely the $f(R)$ expansion history matches that of $\Lambda$CDM.
From the high curvature limit of $f(R)$, 
\begin{eqnarray}
\lim_{\bar{R} \rightarrow \infty} \bar{f}(\bar{R}) \approx - \frac{c_1}{c_2} m^2 + \frac{c_1}{c_2^2} m^2 \left(\frac{m^2}{\bar{R}}\right)^n,
\end{eqnarray}
we see that larger $c_1/c_2^2$ results in closer match to $\Lambda$CDM 
(i.e., $\bar{f}(\bar{R})$ approaches a constant as $c_1/c_2^2 \rightarrow 0$) and smaller $n$ 
mimics $\Lambda$CDM until later in the expansion history.
It is also important to note that the condition $\bar{R}_{0} \gg m^2$ requires that 
$\bar{R} \gg m^2$ at the background level throughout the expansion history 
because $\bar{R}$ is a monotonically decreasing function of time.
Thus, in our numerical simulations, we are able to simplify the field 
equations given the high curvature assumption.

Phenomenologically viable $f(R)$ models require that $|f_{R0}|$ be small 
so as to not significantly alter the cosmic expansion history, limiting
the field strength to $|f_{R0}| \sim 1$ with current observations \citep{song07b}
and to $|f_{R0}| \sim 0.01$ with future weak lensing surveys \citep{hu07a}.
An even stronger constraint can be placed on the field by considering 
the dynamics of our galaxy, resulting in an upper bound of $|f_{R0}| \sim 10^{-6}$
\citep{hu07a}.
However, such analyses do not self-consistently model large scale clustering
that potentially relaxes the tight upper bound.

In the high curvature regime, our $f(R)$ model can be expanded as,
\begin{eqnarray}
f_{R} &=&  -\frac{n c_1 (R/m^2)^{n-1}}{(c_2 (R/m^2)^{n} + 1)^2} \nonumber \\
&\approx& -\frac{n c_{1}}{c_{2}^2} \left( \frac{m^2}{R} \right) ^{n+1}, \label{eqn:frdef}
\end{eqnarray}
valid when $R \gg m^2$.
Furthermore, the field equation (\ref{eqn:field}) can be simplified by neglecting
the small $f_R R$ term, thus resulting in
\begin{eqnarray}
3 \Box f_{R} - R - 2f = -\kappa^2 \rho.
\end{eqnarray}
Assuming the Friedmann-Robertson-Walker 
(FRW) metric, we subtract the spatially constant background quantities
and thus the field equation becomes an equation for the perturbations, 
\begin{eqnarray}
3 \Box \delta f_{R} - \delta R = -\kappa^2 \delta \rho,\label{eqn:tdfield}
\end{eqnarray}
where $\delta R \equiv R - \bar{R} \equiv R - R_{\rm \Lambda CDM}$, 
$\delta f_{R} = f_R - f_{R}(\bar{R})$, $\delta \rho = \rho - \bar{\rho}$,
and the term proportional to $\delta f \approx f_{R} \delta R$ has been
 neglected because $f_{R} \ll 1$.
Finally, we work in the quasi-static limit in which $\nabla f_R \gg \partial f_R / \partial t$,
and the field equation is reduced to 
\begin{eqnarray}
\nabla^2 f_{R} = \frac{1}{3}\left[\delta R(f_{R}) - 8 \pi G \delta \rho\right]. \label{eqn:frorig}
\end{eqnarray}
The $f_R$ field is coupled to the gravitational potential, $\phi$, via
\begin{eqnarray}
\nabla^2 \phi = \frac{16 \pi G}{3} \delta \rho - \frac{1}{6} \delta R(f_R) \label{eqn:potorig},
\end{eqnarray}
for which the derivation can be found in \cite{hu07a}.
Equations (\ref{eqn:frorig}) and (\ref{eqn:potorig}) define a system of equations 
for the gravitational potential, the solution to which can be used to evolve matter
particles.
Note that in the limit of ordinary GR, $\delta R = 8 \pi G \delta \rho$, and the 
equation for the gravitational potential reduces to the unmodified equation,
\begin{eqnarray}
\nabla^2 \phi = 4 \pi G \delta \rho.
\end{eqnarray}
The strategy that we use to simulate $f(R)$ cosmology boils down to solving
the two field equations, Eq.~\ref{eqn:frorig} and \ref{eqn:potorig}, and using the resulting peculiar potential to update
the dark matter density.

%%%%%%%%%%%%%%%%%%%%%%%%%%%%%%%%%%%%%%%%%
\section{Numerical Methods} \label{sec:numerical}
%%%%%%%%%%%%%%%%%%%%%%%%%%%%%%%%%%%%%%%%%
Our N-body simulation is based on the Particle-Mesh algorithm (hereafter PM), in which the 
particle densities are interpolated onto a regular grid and the potential
is solved only at the grid points
\citep[see, e.g.,][]{hockney81,bertschinger98a}.
The choice to use a field representation of the potential instead of particle
pair representation (as in Tree codes) is motivated by the fact that 
the change in the inter-particle forces are mediated by an auxiliary scalar field, 
namely $f_{R}$.
The force law between a pair of particles in the 
presence of $f_{R}$ depends on the entire behavior of the field between the two
particles and thus cannot be expressed as a simple function of their separation.
In this section, we outline the method we developed to obtain $\phi$ by
numerically solving equations (\ref{eqn:frorig}) and (\ref{eqn:potorig}).

%-----------------------------------------------------------------------------------------------------
\subsection{Code Units}
%-----------------------------------------------------------------------------------------------------
The code units are based on the convention used in \cite{shandarin80a,kravtsov97a},
where
\begin{eqnarray}
\tilde{r} = a^{-1} \frac{r}{r_0},\ \ \tilde{t} = \frac{t}{t_0},\ \ \tilde{p} = a \frac{v}{v_0},\nonumber  \\
\tilde{\phi} = \frac{\phi}{\phi_0},\ \ \tilde{\rho} = a^3 \frac{\rho}{\rho_0},\ \ \tilde{R} = a^3 \frac{R}{R_0}, \nonumber
\end{eqnarray}
and
\begin{eqnarray}
r_{0} &=& \frac{L_{\rm box}}{N_{\rm g}}, \\
t_{0} &=& H_{0}^{-1}, \\
v_{0} &=& \frac{r_{0}}{t_{0}}, \\
\rho_{0} &=& \rho_{\rm c, 0} \OmegaMnow, \\
\phi_{0} &=& v_{0}^{2}, \\
R_{0} &=& m_0^2 = \frac{8 \pi G \rho_{0}}{3}.
\end{eqnarray}
In the above definitions, $L_{\rm box}$ is the comoving simulation box size in $h^{-1}$ Mpc, 
$N_{\rm g}$ is the number of grid cells in each direction, $H_{0}$ is the Hubble parameter
today, $\rho_{\rm c,0}$ is the critical density today, and $\Omega_{\rm M,0}$ is the fraction
of non-relativistic matter today relative to the critical density.

Before converting the field equations to code units, we restore the factors of
$c$ in Eqn (\ref{eqn:frorig}), which results in
\begin{eqnarray}
\nabla^2 f_{R} = \frac{1}{3 c^2}\left[\delta R(f_{R}) - 8 \pi G \delta \rho\right]. \label{eqn:fc1}
\end{eqnarray}
Substituting all physical quantities and derivatives in Eqn (\ref{eqn:fc1}) by
their code unit equivalent yields  
\begin{eqnarray}
\tilde{\nabla}^2 \delta f_{R} &=& \frac{\Omega_{\rm M,0}}{a \tilde{c}^2} \left[ \frac{\delta \tilde{R}}{3}  - \delta \right] \label{eqn:frcode},
\end{eqnarray}
where 
\begin{eqnarray}
\delta = \frac{\rho - \bar{\rho}}{\bar{\rho}}.
\end{eqnarray}
The quantity  $\tilde{c}$ is the speed of light in code units, i.e., $\tilde{c} = c / v_0$.
Similarly, we transform Eq.~(\ref{eqn:potorig}) into code units and obtain
\begin{eqnarray}
\tilde{\nabla}^2 \tilde{\phi} &=& \frac{\Omega_{\rm M,0}}{a} \left[2 \delta - \frac{1}{6} \delta \tilde{R} \right] \label{eqn:potcode}.
\end{eqnarray}
From this point on, we work in code units unless otherwise specified, and we
drop the tilde from code variables.

%-----------------------------------------------------------------------------------------------------
\subsection{Solving for $f_{R}$ and $\phi$}
%-----------------------------------------------------------------------------------------------------
The field equation for $f_{R}$ is a non-linear elliptical partial differential equation
because the function $R(f_{R})$ is non-linear in general.
Because of this non-linearity, standard spectral methods (e.g., FFT method)
are not applicable and we therefore resort to a 
non-linear relaxation method for the solution.
Given a general partial differential equation,
\begin{eqnarray}
L(u) &=& f, \label{eqn:basic}
\end{eqnarray}
where $L(u)$ is a non-linear differential operator on the field $u$ and $f$ is
the source term, let us define 
\begin{eqnarray}
\lsc(u) &\equiv& L(u) - f.
\end{eqnarray}
The problem of finding the solution, $u$, for Eq.~(\ref{eqn:basic}) becomes a problem of 
finding the root of $\lsc(u)$.
The root-finding is carried out by an iterative Newton-Raphson algorithm, 
which defines the relaxation method \citep[see, e.g., ][]{press92a}, 
\begin{eqnarray}
u^{\rm new} &=& u^{\rm old} - \frac{\lsc(u^{\rm old})}{\partial \lsc / \partial u^{\rm old} } .
\end{eqnarray}
The only remaining difficulty is determining the appropriate form of $\lsc(u)$ and 
its discretization.
For our model of $f(R)$, the differential operator $\lsc(f_R)$ is defined as
\begin{eqnarray}
\lsc(f_R) = \frac{a c^2}{\Omega_{\rm M,0}} \nabla^2 f_R - \frac{R(f_R) - \bar{R}}{3} + \delta. \label{eqn:difforig}
\end{eqnarray}

A naive and straightforward discretization of Eq.~\ref{eqn:frcode} suffers from
severe numerical instability.
The differential operator in Eq.~\ref{eqn:difforig} is not defined for non-negative values 
of $f_R$ (see Eq.~\ref{eqn:frdef}), and thus the iterative solution fails as 
soon as an iteration oversteps the true solution into the forbidden region.
We find that placing an artificial ceiling on the value of $f_R$ does not adequately
solve the problem because the stability becomes conditional on very precise tuning
of the ceiling due to the steepness of $R(f_R)$ near $f_R = 0$.
The best way to avoid the problem in this model is to redefine the field as
\begin{eqnarray}
f_R = \bar{f_R} e^u
\end{eqnarray}
and solve for $u$ instead of $f_R$.  
With this substitution, the new field $u$ has infinite domain and does not suffer from
abrupt instabilities.
The differential operator thus becomes
\begin{eqnarray}
\lsc(f_R) = \frac{a c^2}{\Omega_{\rm M,0}} \bar{f_R} \nabla \cdot (e^u \nabla u) - \frac{R(\bar{f_R} e^u) - \bar{R}}{3} + \delta. \label{eqn:diffu}
\end{eqnarray}
In our implementation, the differential operator is discretized as a variable coefficient
Poisson equation, in which the value of $u$ at a grid cell identified by $i,j$, and $k$
is updated using the red-black Newton-Gauss-Seidel (NGS) scheme \citep{press92a}.
The actual discretization is too long to be displayed in the body of the text, and 
is instead placed in Appendix \ref{appendix:discretization}.
We use periodic boundary conditions to close the NGS scheme at the edges of our
simulation boxes.

One potential problem with this substitution is the fact that we added additional
non-linearity into a problem that is already highly non-linear.
In particular, the additional non-linearity is exponential, and thus the errors
in the discretized solution, $u_{i,j,k}$, are potentially enhanced exponentially.
In our experience running cosmological simulations with $f(R)$ modification,
we find that the effects of this addition are negligible in all cases and the iterations
converge to the same solution whether this substitution is made or not, while
avoiding the singularity of the curvature field at $f_R = 0$.

Given the solution to $f_R$ using the NGS algorithm and the density field,
we can compute the solution to $\phi$ using Eq.~\ref{eqn:potcode}.
The equation is a standard linear Poisson equation and thus is solved
efficiently using the fast Fourier transform method.
We adopt the cloud-in-cell density interpolation algorithm (hereafter CIC), which treats a 
particle as a uniform density cube that is equal in size to the underlying grid cell,
to compute the density field.

%-----------------------------------------------------------------------------------------------------
\subsection{Multigrid Acceleration}
%-----------------------------------------------------------------------------------------------------
In general, relaxation methods for solutions of partial differential equations
are very inefficient and computationally expensive.
However, this inefficiency can be mostly alleviated by the use of multigrid
techniques.
The multigrid method \citep{brandt73} is a class of algorithms to solve partial 
differential equations using a hierarchy of discretizations with increasingly
coarse grids.
In particular, it exploits the fact that a typical relaxation schemes, such
as the NGS scheme previously detailed, smooth out the
high frequency residual errors faster than the low frequency error modes.
Thus, by interpolating the low frequency error modes onto a coarser grid,
the same error mode now appears as a high frequency mode on the
coarser grid and can be efficiently reduced by relaxation passes on that
grid.

In our working code, we implement a particular type of multigrid method
called the Full Approximation Scheme \citep[][hereafter FAS]{briggs00a},
which is particularly suited for non-linear boundary value partial differential
equations.
In order to define the FAS algorithm, we first define two grids,
the fine grid ($\Omega_{h}$), and the coarse grid ($\Omega_{2h}$).
In addition, we define two inter-grid
operators, one that restricts a fine grid field onto a coarse grid, and one 
that interpolates a coarse grid field onto a fine grid.  
We label the former as $I_{h}^{2h}$ and the latter as $I_{2h}^{h}$, and
leave the details of their implementations to Appendix \ref{appendix:multigrid}.
Finally, we denote the quantities defined on a coarse grid by a superscript $2h$
and the quantities defined on a fine grid by a superscript $h$.

The two grid FAS scheme can be summarized by the following.  
Given the initial guess to the solution, $u^{h}$, the FAS algorithm
can be summarized as follows: 
\begin{eqnarray}
\bullet && \text{Relax on } \Omega_{h} \text{(pre-smoothing)}\nonumber \\
\bullet && \text{Compute: } f^{2h} = I_{h}^{2h}(f^{h} - L^{h}(u^h)) + L^{2h}(I_h^{2h} u^h) \nonumber \\
\bullet && \text{Solve: } L^{2h}(u^{2h}) = f^{2h} \nonumber \\
\bullet && \text{Correct the solution: } u^h_{new} = u^h + I_{2h}^h(u^{2h}-I_h^{2h} u^h) \nonumber \\
\bullet && \text{Relax on } \Omega_{h} \text{(post-smoothing)} \nonumber
\end{eqnarray}
In a true multigrid method, the third step above is also carried out using 
a multigrid method, thus forming the recursive hierarchy of coarser grids.
The recursion is stopped when the number of coarse grid cells per dimension 
reaches four; the solution at this level is obtained by iterating NSG until
convergence.
All relaxations are carried out using the NGS scheme that is appropriate for
the given coarse (or fine) grid.

In our implementation, we use the full-weighted restriction operator ($I_{h}^{2h}$)
and its adjoint, the bilinear interpolation as the prolongation operator ($I_{2h}^{h}$).
In practice, we find that the choice of the restriction and prolongation operators
does not affect the stability of convergence for most typical cosmological simulation.
However, we find the convergence rate when using full-weighting and bilinear interpolation
to be better than other, simpler choices.

%-----------------------------------------------------------------------------------------------------
\subsection{Particle Dynamics} \label{subsection:pdynamics}
%-----------------------------------------------------------------------------------------------------
Given the solution for $f_R$ field, the Newtonian potential is computed by solving Eqn (\ref{eqn:potcode}), 
which is a linear Poisson equation with a non-standard (i.e., non-GR) source term.
Because of its linearity, such equation can be solved efficiently using a fast Fourier transform method
\citep[FFT, see][]{hockney81}.
The forces on the dark matter particles are computed
using Newton's equations in comoving code coordinates,
\begin{eqnarray}
\frac{d x}{d a} &=& \frac{p}{\dot{a} a^2}, \\
\frac{d p}{d a} &=& - \frac{\nabla \phi}{\dot{a}}.
\end{eqnarray}
Because the expansion history of our $f(R)$ model is tuned to that of flat $\Lambda$CDM, 
$\dot{a}$ is computed simply as 
\begin{eqnarray}
\dot{a} = a^{-1/2} \sqrt{\Omega_{\rm M,0} + \Omega_{\Lambda,0} a^3},
\end{eqnarray}
where the over-dot denotes a derivative with respect to coordinate time in code units.
The gradient of the gravitational potential is computed using a standard finite difference
scheme consistent with our grid density assignment operator, namely the CIC scheme.
We evolve the particles using a second-order accurate leap-frog method \citep{klypin97a}.

%%%%%%%%%%%%%%%%%%%%%%%%%%%%%%%%%%%%%%%%%
\section{Tests of the code} \label{sec:tests}
%%%%%%%%%%%%%%%%%%%%%%%%%%%%%%%%%%%%%%%%%

Our N-body simulation is implemented in C++ with shared memory parallelization
using OpenMP.
In the following subsections, we present various tests of the implementation 
to assess the correctness of the code as well as its computational speed.

%-----------------------------------------------------------------------------------------------------
\subsection{Analytically tractable case} \label{subsection:atc}
%-----------------------------------------------------------------------------------------------------
\begin{figure*}
    \begin{center}
      \resizebox{83mm}{!}{\includegraphics[angle=0]{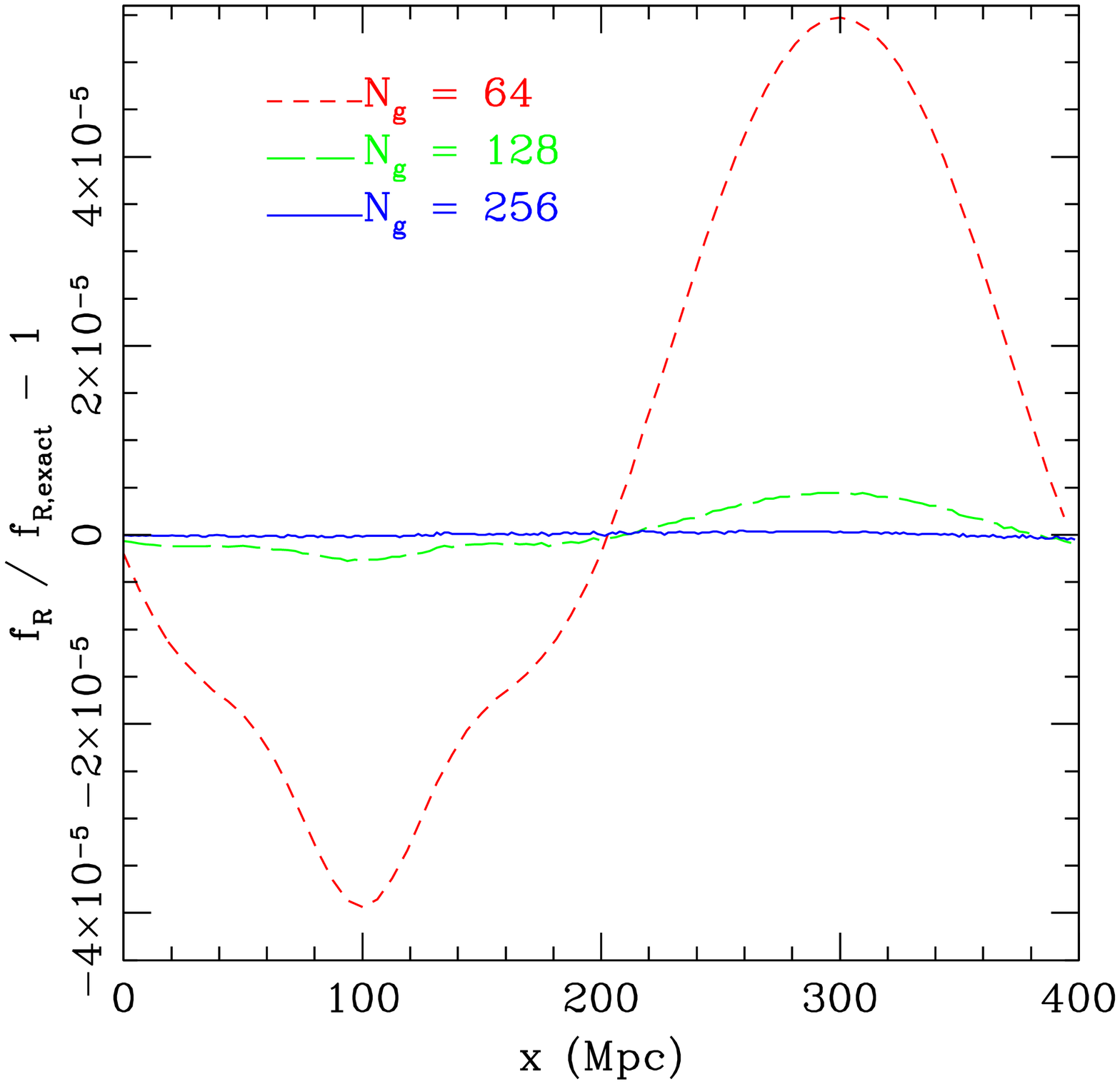}}
      \resizebox{83mm}{!}{\includegraphics[angle=0]{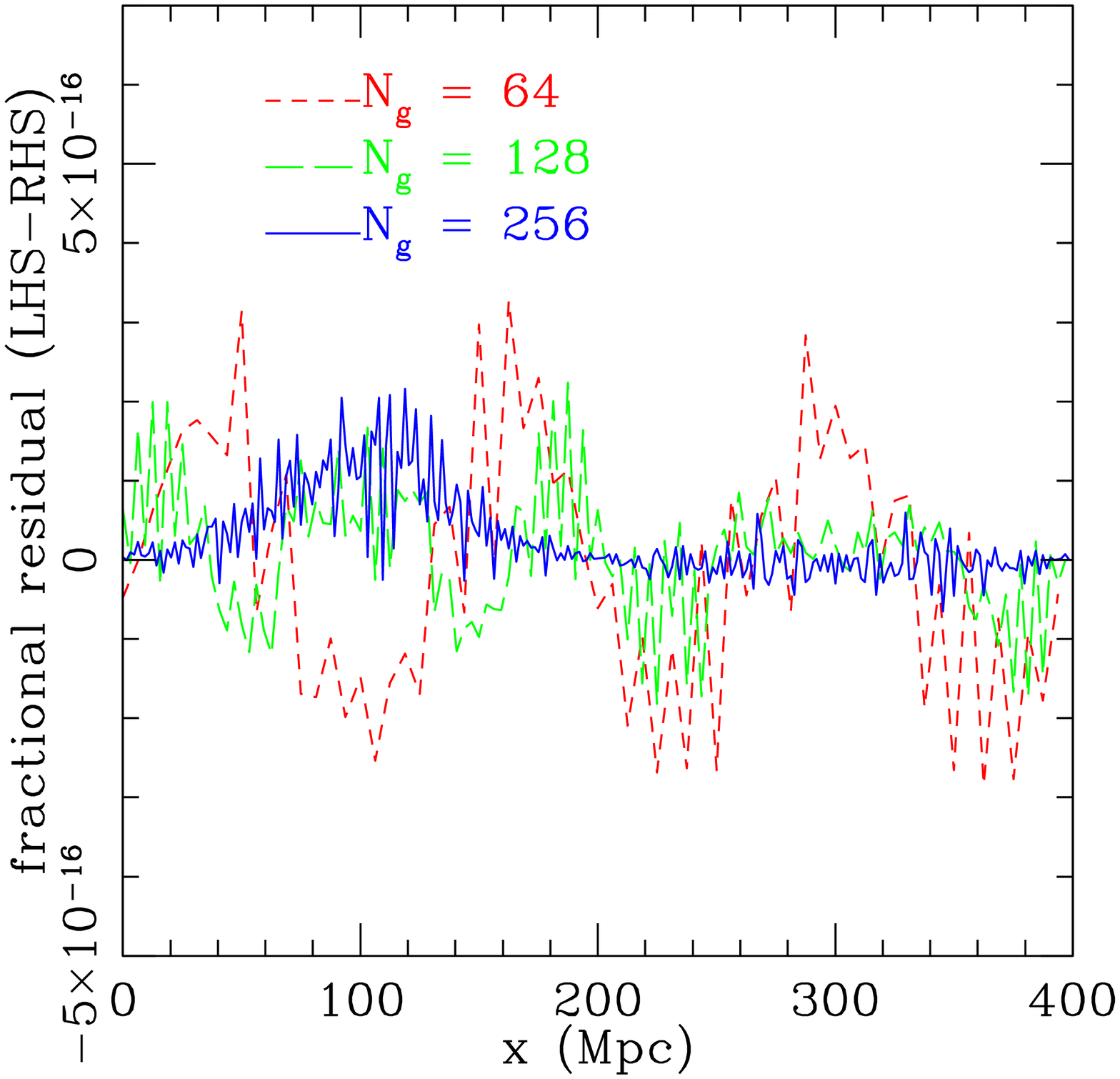}}
    \end{center}
  \caption{ The fractional discretization error ({\it left}), shown as $f_{R, {\rm code}} / f_{R, {\rm exact}} - 1$, and 
	the fractional residual error ({\it right}), $\lsc(f_R)$,
	for the analytically tractable case of \S \ref{subsection:atc}.
	The discretization error for the highest resolution box is $\sim 10^{-6}$,
	and allows us to define a sensible stopping criteria for the multigrid
	iterations.
	The fractional residual error is floored by limits on numerical precision,
	which is $\sim 10^{-15}$.
	} 
  \label{plot:analytical}
\end{figure*}

First, we construct a density field $\delta$ that has a known $f_R$ solution
to check that the code can recover the true solution.
In 1D, the field equation (\ref{eqn:frcode}) becomes
\begin{eqnarray}
\frac{\partial^2 f_R}{\partial^2 x} = \frac{\Omega_{\rm M,0}}{a c^2} \left( \frac{\delta R}{3} - \delta\right),
\end{eqnarray}
which, after setting all constants to 1 and taking $n = 1$ and $\bar{R} = 1$, is
\begin{eqnarray}
\frac{\partial^2 f_R}{\partial^2 x} = \frac{1}{3} \left[ \left(-f_R\right)^{-1/2} - 1 - 3 \delta\right] \label{eqn:1d}.
\end{eqnarray}
We take the analytic solution to be
\begin{eqnarray}
f_R = \sin\left(\frac{2 \pi x}{L_{\rm box}}\right) - 2 \label{eqn:extsolution},
\end{eqnarray}
and substitute the solution into equation (\ref{eqn:1d}) to find 
\begin{eqnarray}
\delta &=& \left(\frac{2 \pi}{L_{\rm box}}\right)^2 \sin \left(\frac{2 \pi x}{L_{\rm box}}\right) \nonumber \\
&&+ \frac{1}{3}\left[ 2 - \sin \left(\frac{2 \pi x}{L_{\rm box}}\right)\right]^{-1/2} - \frac{1}{3}. \label{eqn:1dden}
\end{eqnarray}
Thus, setting the density field to the expression in (\ref{eqn:1dden}) and numerically
solving for $f_R$ allows us to compare the code accuracy against the known solution,
Eq.~\ref{eqn:extsolution}.

In Fig.~\ref{plot:analytical}, we show the results of applying our code to this
problem with three different grid resolutions.
In the left panel, fractional discretization errors, defined as $f_{R, {\rm code}} / f_{R, {\rm exact}} - 1$,
are shown, and the residual errors, defined as $\lsc(u)$, are shown in the right panel.
The average discretization error decreases by approximately an order of magnitude
for each two-fold increase in grid resolution.
Thus, for $N_{\rm g} = 512$, we expect the discretization error to be on the order
of $10^{-7}$, and therefore gives us a useful stopping criteria for solution
convergence based on the requirement that the discretization error is the dominant
source of error.
We terminate the iterative multigrid process when the residual error (which can be
computed without knowing the true solution) is smaller than $10^{-10}$, guaranteeing
the residual error to be much smaller than discretization error for our highest 
resolution simulations.

%-----------------------------------------------------------------------------------------------------
\subsection{Point mass} \label{subsection:npm}
%-----------------------------------------------------------------------------------------------------
\begin{figure}
  \begin{center}
  \resizebox{84mm}{!}{\includegraphics[angle=0]{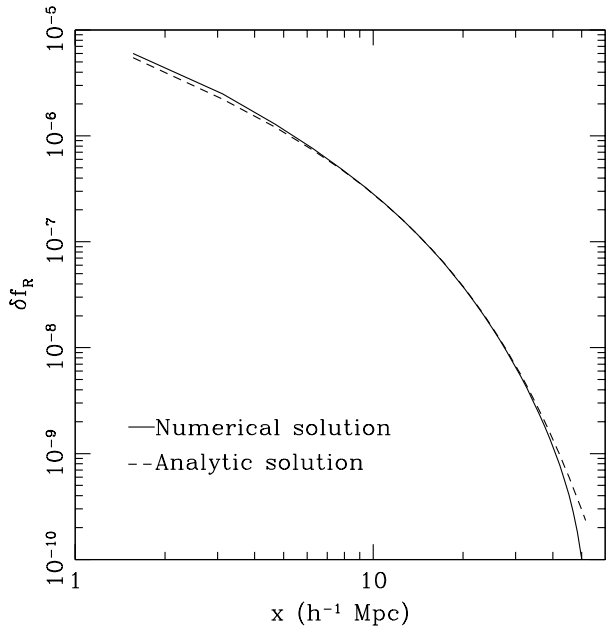}}
  \end{center}
  \caption{The computed and expected solutions for the point mass
  test. 
  The analytic solution is arbitrarily normalized such that it and the numerical
  solutions match at scales where the simulation is expected to best match the
  analytic solution.  
  This scale corresponds to $x = 10 h^{-1}$ Mpc.
  The background effective field mass is $0.13 h$ Mpc$^{-1}$.
  } 
  \label{plot:npm1}
\end{figure}

In this test, we place a point mass (consistent with the CIC scheme) at the 
origin of a periodic simulation box, with the corresponding density field given 
by
\begin{eqnarray}
\delta^{(1)}_{i,j,k} &=& \left\{ \begin{array}{rl} 10^{-4} N_g^3 & \mbox{if $i=j=k=0$} \\ \\ -10^{-4} & \mbox{otherwise} \end{array} \right.
 \end{eqnarray}
 where the indices $i$, $j$, and $k$, refer to the grid cell indices in $x$, $y$, and
 $z$ directions, respectively.
In the subsequent discussion in this section, we restore the tilde notation to distinguish
code variables and physical variables. 
With this density configuration, outside the grid cell at the origin, we expect that
\begin{eqnarray}
\tilde{\nabla}^2 \delta f_R &\approx& \frac{\Omega_{\rm M,0}}{a \tilde{c}^2} \delta \tilde{R} \nonumber \\
&\approx& 3 \frac{\Omega_{\rm M,0}}{a \tilde{c}^2} m_{\rm eff}^2 \delta f_R \label{eqn:fryukawa}
\end{eqnarray}
where $m_{\rm eff}$ is the effective field mass of the $f_R$ field, given in physical
coordinates by 
\begin{eqnarray}
m_{\rm eff}^2 = \frac{1}{3}\left(\frac{1+f_R}{f_{RR}}-R\right) \approx \frac{1}{3} f_{RR}^{-1}
\end{eqnarray}
for small $f_R$.
The quantity $f_{RR}$ is the second derivative of $f(R)$ with respect to $R$.
Thus, the solution to Eq.~(\ref{eqn:fryukawa}) is a Yukawa like
potential,
\begin{eqnarray}
\delta f_R \propto \frac{e^{-\tilde{m}_{\rm eff} r}}{r},
\end{eqnarray}
where
\begin{eqnarray}
\tilde{m}_{\rm eff} = \sqrt{3 \frac{\Omega_{\rm M,0}}{a \tilde{c}^2}} m_{\rm eff}
\end{eqnarray}
is the effective comoving field mass in code units.

Fig.~\ref{plot:npm1} shows the $f_R$ field solution computed on a $N_{\rm g}
= 256$ grid.
The simulation parameters were: $L_{\rm box} = 400$ Mpc, $\Omega_{M,0} = 0.3$,
$\Omega_{\Lambda,0} = 0.7$, $h = 0.7$, and $f_{R,0} = -10^{-5}$.
On the same plot, the analytic solution with the expected $\tilde{m}_{\rm eff}$ is plotted.
%The two solutions are not expected to match exactly because of several reasons.
%First, the Yukawa solution assumes a point mass, but the density field in this
%test is constrained by computation resolution and therefore does not represent
%a true point mass.
The $\sim 10\%$ discrepancy at $x \sim 2 h^{-1}$ Mpc is due to the fact that
 the approximation $\delta R \approx m^2_{\rm eff} \delta f_{R}$
is only valid for small $\delta f_R$ and thus is only appropriate at large distances
away from the central near point mass. 
Hence, the Yukawa solution in Fig.~\ref{plot:npm1} is normalized such that the solution
best matches the simulation at $x \sim 10\ h^{-1}$ Mpc where the $\delta f_R$ becomes
sufficiently small.
At $x \gtrsim 40 h^{-1}$ Mpc, the solution $\delta f_R$ approaches the level of discretization
error for $N_g = 256$ ($\sim 10^{-6}$) and hence begins to show deviations from the analytic solution.
%Finally, the computation box uses periodic boundary conditions appropriate for
%general cosmological simulations, and thus does not represent a true isolated
%point mass.
%The periodic boundary condition is responsible for the mismatch to the Yukawa
%solution at $x \sim 50\ h^{-1}$ Mpc.

%-----------------------------------------------------------------------------------------------------
\subsection{Dynamics} \label{subsection:dynamics}
%-----------------------------------------------------------------------------------------------------
\begin{figure}
  \begin{center}
  \resizebox{84mm}{!}{\includegraphics[angle=0]{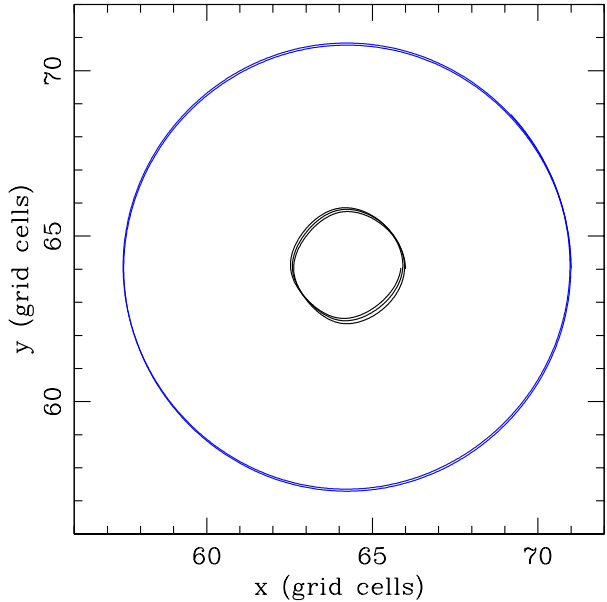}}
  \end{center}
  \caption{Orbit test. A point mass is placed at the center of a $128^3$ grid, and
  the trajectory of test particles are simulated without $f(R)$ modification.
  The inner particle is placed at two grid cells away from the center, while the outer
  particle is placed seven grid cells away.  Both particles are given initial velocities
  such that in they would be in circular orbit without $f(R)$ modifications.  
   } 
  \label{plot:dyn1}
\end{figure}
\begin{figure}
  \begin{center}
  \resizebox{84mm}{!}{\includegraphics[angle=0]{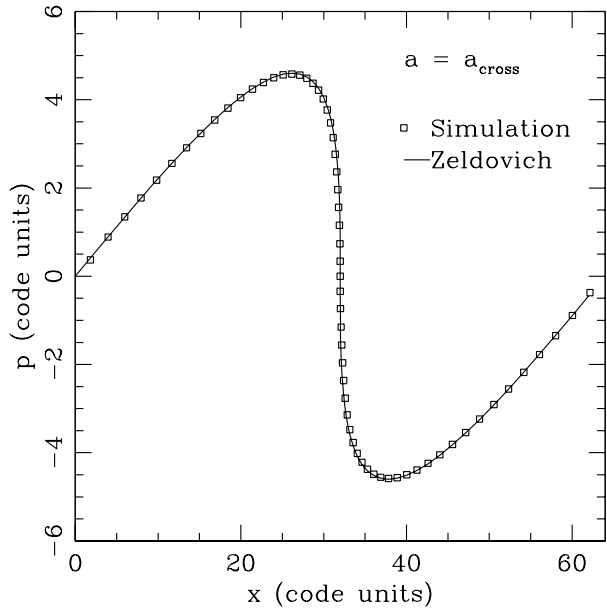}}
  \end{center}
  \caption{Zeldovich 1D plane wave collapse with no $f(R)$ modifications.  
  Particle momenta ($p$) is plotted against particle positions ($x$).
  The points represent the simulation output, and the line represents the exact solution
  given by the Zeldovich approximation.  The snapshot is taken at $a = a_{\rm cross} = 1$.
   }
  \label{plot:zd1}
\end{figure}
\begin{figure}
  \begin{center}
  \resizebox{84mm}{!}{\includegraphics[angle=0]{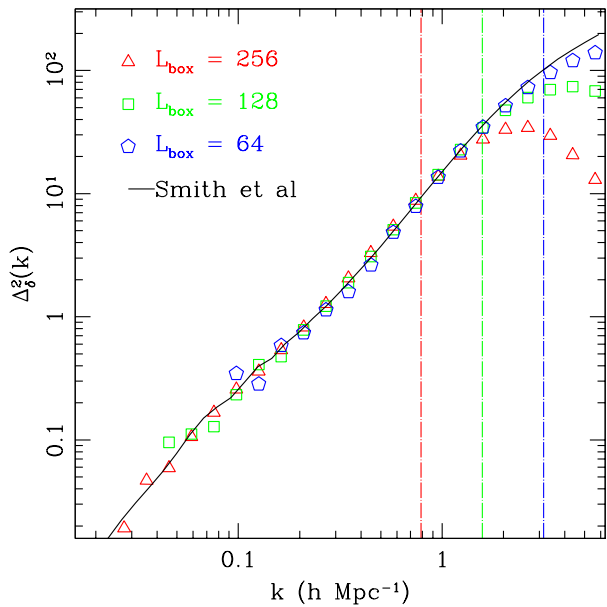}}
  \end{center}
  \caption{Power spectra computed from cosmological simulations without
  $f(R)$ modifications.
  The Smith et. al. fit is plotted solid line, showing good agreement
  with our simulations.  The vertical lines, from left to right, are the particle half-Nyquist
  scales, defined as $k_{\rm hN} = \pi N_{p} / (4 L_{\rm box})$, 
   for $L_{\rm box}=256$, $128$, and $64$ $h^{-1}$ Mpc boxes.} 
  \label{plot:pspec.gr}
\end{figure}

As previously described in \S \ref{subsection:pdynamics}, the particle trajectories
are integrated using a second-order accurate leap-frog scheme.
In Fig.~\ref{plot:dyn1}, we show the orbits of test particles orbiting around a 
central point mass.
The test particles are placed at two and seven grid cells away from the point mass
and are given initial velocity such that they would be in circular orbits if there were 
no integration errors and no $f(R)$ modifications.
The solid lines show the simulated orbit for the two different radii.
Far away from the point source, where the forces are adequately resolved, the
orbit is nearly circular as expected for unmodified GR.
%The same orbit under $f(R)$ modification is not circular because the two body
%force law is different under such modification.
%In this setup, the two body forces are enhanced by the $f_R$ field, and thus
%the resulting orbit is tighter than normal.
The smaller orbit, with a theoretical radius of two grid cells, is not well resolved
by the underlying potential grid.
However, the simulated orbit is still nearly circular, with only a few percent
deviation after two complete orbits.
The two-body force law, and its modifications due to $f(R)$ models,
 will be further studied in \S \ref{subsection:tbf}.

Another popular test of the particles dynamics as well as of the accuracy of 
the potential solver is the Zeldovich pancake collapse test \citep{klypin83a,efstathiou85a,kravtsov97a}.
The evolution of a 1D sine wave is described exactly by the Zeldovich solution,
\begin{eqnarray}
x_{i}(a) &=& q_{i} + A G(a) \sin\left(\frac{2 \pi q_{i}}{L_{\rm box}}\right) \\
p_{i}(a) &=& A a^2 \dot{G}(a-\Delta a/2) \sin\left(\frac{2 \pi q_{i}}{L_{\rm box}}\right)
\end{eqnarray}
where $q_{i} = i \Delta x$, $G(a)$ is the linear growth function, $\Delta a$ is the 
time step used in the simulation, and $A$ is a normalization constant that fixes
the time at shell crossing.
By setting the initial conditions at $a = a_{\rm init}$ according to the Zeldovich
solution, we can compare the subsequent particle evolution to the same
solution to gauge the accuracy of our particle dynamics.

For our test, we use $\Omega_{\rm M,0} = 0.24$, $\Omega_{\rm \Lambda,0} = 0.76$,
$h = 0.73$, and no $f(R)$ modifications.
The simulation box is set to $L_{\rm box} = 100$ $h^{-1}$ Mpc, and we use
$64^{3}$ particles and $64^{3}$ grid cells.
Fig.~\ref{plot:zd1} shows the particle momenta versus the particle positions
at the time of shell crossing, which is set to be $a=1$.
The dots denote the simulation output and the line represents the exact solution
given by the Zeldovich approximation.
As expected, the agreement between our code and the Zeldovich solution is 
good, with maximum deviation less than 1\%.

As another check of the particle dynamics, we have run several cosmological
simulations without $f(R)$ modifications to check our results against the 
Smith et. al. fits to the power spectrum \citep{smith03a}.
We run three simulations with box sizes ($L_{\rm box}$) of 256 $h^{-1}$ Mpc (L256),
128 $h^{-1}$ Mpc (L128), and 64 $h^{-1}$ Mpc (L64).
All simulations have the following cosmological parameters resembling the 
third year Wilkinson Microwave Anisotropy Probe \citep[][WMAP3]{spergel07a} results: $\Omega_{\rm M,0} = 0.24$, $\Omega_{\Lambda,0}
= 0.76$, $\Omega_{\rm b, 0} = 0.04181$, $H_{0} = 72$ km/s/Mpc, and $\sigma_{8} = 0.76$.
We take the slope of the primordial power spectrum, $n$, to be 0.958.

The initial conditions for the simulations are created using Enzo \citep{oshea04a},
a publicly available cosmological N-body + hydrodynamics code.
Enzo uses the Zeldovich approximation to displace particles on a uniform grid according
to a given initial power spectrum.
We use the initial power spectra given by Eistenstein \& Hu \citep{eisenstein98a},
not including the effects of baryon acoustic oscillations.
The simulations are started at $z = 49$.

All simulations are run with 512 grid cells in each direction (i.e., $N_{\rm g} = 512$) and
with $256^3$ particles.  
Thus, the formal spatial resolutions of the simulations are 0.5 $h^{-1}$ Mpc, 0.25 $h^{-1}$
Mpc, and 0.125 $h^{-1}$ Mpc for the largest, middle, and smallest boxes, respectively.
The corresponding mass resolutions are $2.76\times 10^{11}\ \rm M_{\odot}$,
$3.45\times 10^{10}\ \rm M_{\odot}$, and $4.31\times 10^{9}\ \rm M_{\odot}$.

For each simulation box configuration, we run multiple simulations with different
realization of the initial power spectrum in order to reduce finite sample variance.
Two L256 simulations, three L128 simulations, and four L64 simulations were
run. 

In Fig.~\ref{plot:pspec.gr}, we show the power spectrum of our simulations 
as well as the Smith et. al. fit.
Multiple simulations of the same box size are averaged in the figure.
The vertical lines denote the half-Nyquist scale of the particles, plotted as a 
rough guide to show the resolution limits of our simulations.
We see from the plot that our simulations match the Smith et. al. fit to 
within $\sim 10$\% at most scales smaller than the half-Nyquist scale.
The large scatter at low $k$ is seen because only a small number of the
largest wavelength modes can fit in a simulation box.

%%%%%%%%%%%%%%%%%%%%%%%%%%%%%%%%%%%%%%%%%
\subsection{Time derivative} \label{subsection:tdtest}
%%%%%%%%%%%%%%%%%%%%%%%%%%%%%%%%%%%%%%%%%
\begin{figure}
  \begin{center}
  \resizebox{84mm}{!}{\includegraphics[angle=0]{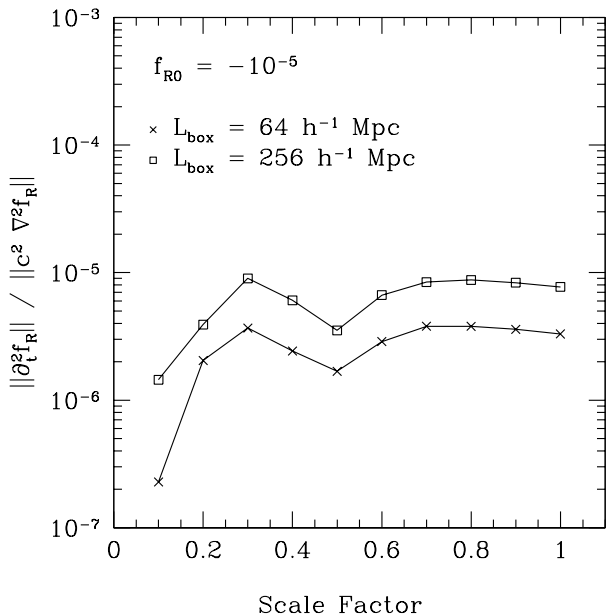}}
  \end{center}
  \caption{The ratio of the time derivative of the $f_R$ field to the spatial
  gradient.  The 2-norm is defined simply as $\sqrt{\langle x^2\rangle}$, where $x$ is either
  $\partial^2_t f_R$ or $\nabla^2 f_R$. The factor of $c^2$ comes from the
  definition of the FRW metric.} 
  \label{plot:tdtest}
\end{figure}

In our simulation, we work in the quasi-static limit of the full field equation 
for $f_R$, Eq.~\ref{eqn:field}, in which we neglect the time derivative term.
There is no simple way to check the validity of our assumption short of 
running a full simulation including the time derivative term; nevertheless,
we can check that our assumption is at least self-consistent by running modified
cosmological simulations with quasi-static assumption and comparing the
time derivative of the $f_R$ field to its spatial derivative.

We use the same simulation setup as the cosmological simulations in \S\ref{subsection:dynamics}
except for the inclusion of $f(R)$ modification with $f_{R0} = -10^{-5}$. 
The time derivative of the field is estimated by finite differencing of fields at successive
time steps.
In Fig.~\ref{plot:tdtest}, we show the fraction of the 2-norms of the time derivative
and spatial derivatives of the field at different scale factors.
Although the ratio varies by time, we see that the time derivative is 
generally much smaller than than the spatial derivative and thus that our
simulations are consistent with our assumptions.

%-----------------------------------------------------------------------------------------------------
\subsection{Code timing and memory requirement}
%-----------------------------------------------------------------------------------------------------
\begin{figure}
  \begin{center}
  \resizebox{84mm}{!}{\includegraphics[angle=0]{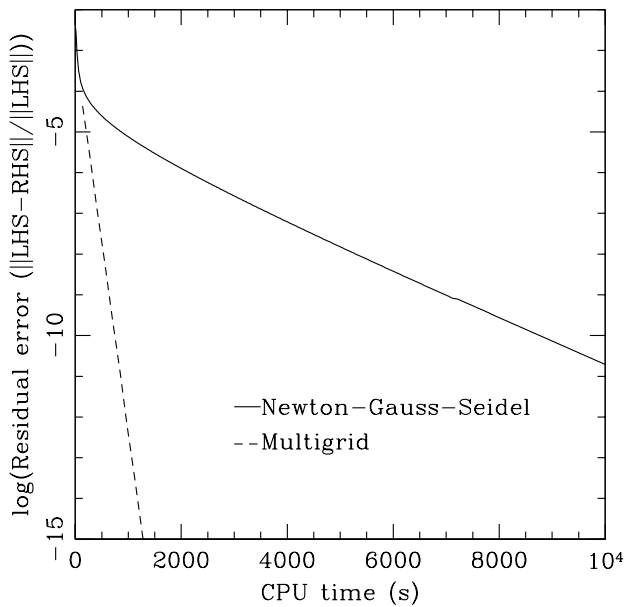}}
  \end{center}
  \caption{The residual error, defined as the 2-norm of the l.h.s minus
  the r.h.s of Eq.~\ref{eqn:fryukawa} divided by the 2-norm of l.h.s,
   is plotted as a function of CPU time spent 
  in the iterative partial differential equations (PDE) solvers.
  The NGS scheme is only efficient at reducing low wavelength error 
  modes and thus the convergence rate stalls as the large wavelength mode
  begins to dominate the error budget.
  The multigrid scheme can reduce error modes of all sizes equally well,
  and thus retains good convergence rate throughout the iterations.
  } 
  \label{plot:npm2}
\end{figure}

Even with the efficiency of the multigrid method, solving a 3D non-linear
partial differential equation with reasonable resolution is still a computationally
expensive endeavor.  
In Fig.~\ref{plot:npm2}, we plot the residual error as a function of the CPU
time spent during the NGS and multigrid iterations of the point mass solution
studied in \S \ref{subsection:npm}.
The computations were carried out on a 2.8 GHz AMD Opteron 2220 processor
without using OpenMP multi-threading (i.e., single threaded).
For the NGS method, the high frequency error modes are reduced more efficiently
than low frequency modes.
Thus, the residual error goes down fast during the initial few hundred seconds of
the NGS iterations, until the low frequency error modes dominate the overall
residual error budget.
Thereafter, the iterations converge slowly, rendering the method unacceptable
for cosmological simulations in which the $f_R$ field must be evaluated 
hundreds of times.
With the multigrid method, because of its use of coarse grids, error modes of
all frequencies decay at nearly the same rate, and thus the overall convergence
is greatly accelerated.
Improved convergence rates using multigrid method, sometimes by as much 
as two orders of magnitudes, makes $f_R$ cosmological simulations a
reality.

As shown in Fig.~\ref{plot:npm2}, convergence to a reasonable error tolerance 
takes $\sim 1000$ s on an AMD Opteron 2220 processor even for a modest 
$N_{\rm g} = 256$ resolution.
The FAS scheme has one of the best computation time scaling, growing 
linearly to the number of grid points (FFT, for comparison, scales as $N \log N$),
and thus a $N_{\rm g} = 512$ simulation will converge in ~10,000 s.
The $f_R$ field equation must be solved for each individual time step of 
the simulation, and thus resulting in potentially thousands of multigrid solutions
totaling a few million seconds of CPU time.
With our discretization scheme, we find that the simulation code spends
$\sim 95\%$ of CPU time solving for the $f_R$ field.
Such percentage is at best a rough estimate, as the number of multigrid
iterations required for convergence depends on the setup of the simulations.

Fortunately, several methods to speed up the simulation are possible.
First, the particle positions between successive time steps does not change
greatly, and thus the $f_R$ solution should not change significantly as well.
The $f_R$ field solution from the previous time step provides a good initial
guess for the iterative solver and we find that it results in $\sim 10\%$ faster
convergence.
Secondly, the FAS scheme is easily parallelized using shared memory 
parallelization with almost perfect scaling with the number of processor cores
available.  
The NGS smoothing used extensively within the FAS scheme can be arranged
in the so called red-black ordering \citep{briggs00a}, such that smoothing
of one particular grid point is independent of all other points except for its
six nearest neighbors.
This locality allows the smoothing to be split into local domains, each of 
which can be computed on separate processors.
Combining such acceleration methods, a typical $N_{\rm g} = 512$
simulation can be carried out in roughly 10 days on our hardware.
We see no difficulties in extending the parallelism to distributed memory,
although we have not implemented such extension.

Memory requirement for the FAS scheme can be high compared to traditional
PM simulations.
The hierarchy of coarser grids add $\sim 15\%$ to the storage requirement
for the $f_R$ field compared to the normal storage requirement for a 
3D grid scaling as $N_{\rm g}^3$.
Furthermore, in order to accelerate the computation, we store the current
residual error and use one additional grid hierarchy of the same size for
transient computations.
For a $N_{\rm g} = 512$ grid, the required memory for our FAS implementation
is $\sim 3.5$ GiB
using double precision floating point format (64 bits each).
For comparison, storage of the density field requires 1 GiB, and the storage of
dark matter particles requires 6 GiB for $512^3$ particles.

%%%%%%%%%%%%%%%%%%%%%%%%%%%%%%%%%%%%%%%%%
\section{Results} \label{sec:results}
%%%%%%%%%%%%%%%%%%%%%%%%%%%%%%%%%%%%%%%%%
In this section, we study a few $f_R$ simulations in order to
highlight the differences between $f_R$ and ordinary $\Lambda$CDM physics.
A more comprehensive study of the non-linear $f_R$ physics is detailed in a
companion paper \citep{oyaizu08b}.

%-----------------------------------------------------------------------------------------------------
\subsection{Two-body force laws} \label{subsection:tbf}
%-----------------------------------------------------------------------------------------------------
\begin{figure}
  \begin{center}
  \resizebox{84mm}{!}{\includegraphics[angle=0]{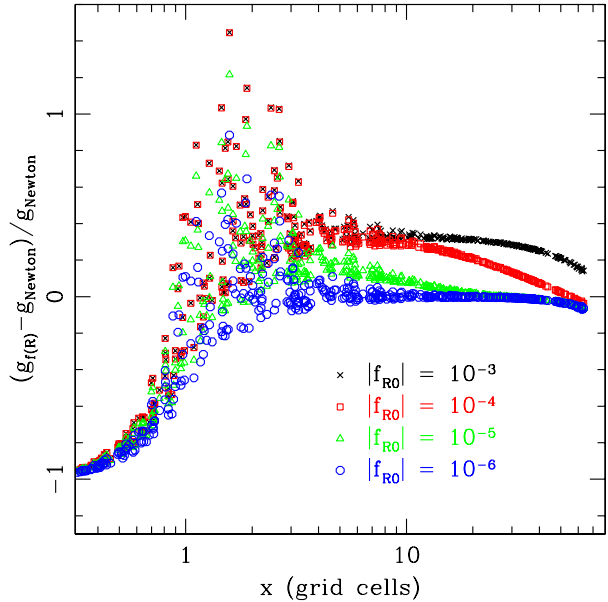}}
  \end{center}
  \caption{Two-body acceleration as a function of separation for several $f_R$ field
  strengths.  
  The accelerations are shown as relative differences from the exact Newtonian
  two-body acceleration.
  The large scatter at $x \sim 2$ is because of truncation errors resulting from
  differencing the potential at scales comparable to the grid size.} 
  \label{plot:tbf1}
\end{figure}
\begin{figure}
  \begin{center}
  \resizebox{84mm}{!}{\includegraphics[angle=0]{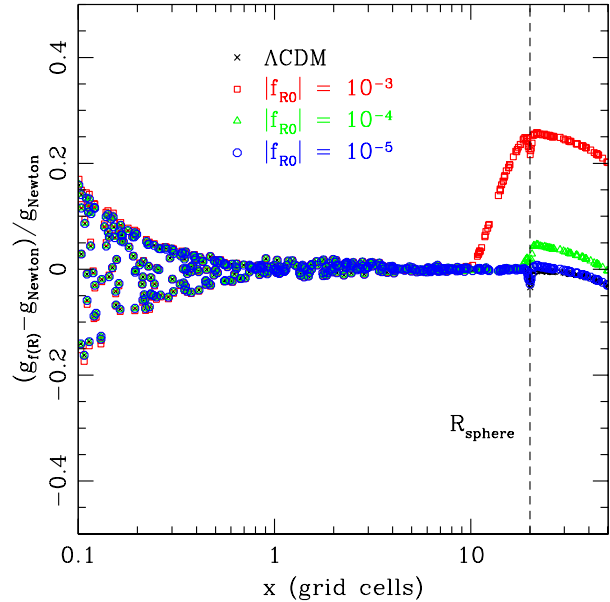}}
  \end{center}
  \caption{Two-body acceleration as a function of separation for several $f_R$ field
  strengths.  The spherical blob has a constant density of $\delta \sim 500$ and
  a radius of 20 grid cells (15.6 $h^{-1}$ Mpc).  
  The increased scatter in the accelerations at the surface of the sphere are due
  to the fact that a smoothness of the spherical surface is limited by the finite
  grid size of the simulation.} 
  \label{plot:tbf2}
\end{figure}
The first interesting case is the study of two body force laws in $f_R$ cosmology.
As is well known, in ordinary GR, two-body forces decay as $r^{-2}$ as the 
separation between the bodies, $r$, is varied.
However, in $f_R$ cosmology, the new field mediates a fifth force that modifies 
two-body interactions and potentially change the large scale structures of the 
Universe.

Such deviations from GR are easy to see using the same setup as the near
point mass text in \S~\ref{subsection:npm}.
We solve for the peculiar gravitational potential, $\phi$, with the $f_{R}$ field and 
compute the gravitational acceleration at random positions
in the simulation box, labeling them {\gfr}.
In addition, we compute the standard Newtonian force using the
inverse square law for the same set of points, labeling them {\gnewton}.
We use a 400 $h^{-1}$ Mpc computation box with $Ng = 256$, $\Omega_M = 0.3$, and $\Omega_{\Lambda} = 0.7$. 
Fig.~\ref{plot:tbf1} shows the relative differences between {\gfr} and {\gnewton} as a function 
of the distance away from the point mass.
For strong $f_R$ fields, the gravitational forces near the central point mass are 
enhanced by as much as $\sim 30\%$ over the unmodified forces.
The range of the force enhancement is determined by the Compton wavelength
of the $f_R$ field, which depends on the local field mass as
\begin{eqnarray}
\lambda_{\rm eff} = m_{\rm eff}^{-1} \sim |f_R| ^ {(n+2)/2(n+1)}. \label{eqn:compton}
\end{eqnarray}
Thus, the stronger the field, the farther out the gravitational enhancement of the 
$f_R$ field propagates, which is confirmed by Fig.~\ref{plot:tbf1}.
For the very weak field with $f_R = -10^{-6}$, the Compton wavelength is much smaller
than the computational grid cells and therefore the gravitationally enhanced regions
are not resolved.

In the strong field limit, the large Compton wavelength of the $f_R$ field makes
it smooth and stiff, resisting the fluctuations in the underlying density field.
Thus, in this limit, the deviation of the field, and hence of the curvature field $R$,
vanishes and we have $\delta R \approx 0$.
The equation for the gravitational potential, Eq.~\ref{eqn:potorig}, reduces to
\begin{eqnarray}
\nabla^2 \phi &=& \frac{16 \pi G}{3} \delta \rho = \frac{4 \pi}{3} \left(\frac{4}{3}G\right) \delta \rho.
\end{eqnarray}
We see that the result is an effective enhancement of the Newton's constant
by 33\%, and hence the two-body forces enhanced by the same amount.
Our two-body numerical solution captures this effect well, with the $f_{R0} = -10^{-3}$
acceleration curve plateauing at the correct value.

Features not due to the presence of the $f_R$ field are also seen in Fig.~\ref{plot:tbf1}.
The large scatter near $x \sim 2$ are due to the lack of 
force resolution at scales comparable to the underlying grid cells.
At scales even smaller than the grid cells, computed forces vanish;
this force decay is essential for the N-body system to remain collision-free.
On the other end of the scale, the periodic boundary condition begins to affect
the forces as the test masses becomes gravitationally attracted to the neighboring
point mass.

At first, Fig.~\ref{plot:tbf1} seems to contradict the stringent solar system tests
carried out so far and would constrain $f_R$ to very small values.
However, the particular $f_R$ model chosen in this study shows the so called
chameleon behavior \citep{khoury04a,khoury04b}, in which the field strength is
pushed to near zero in high density regions.
Small field strength in the high density region causes the effective field mass
to increase and therefore the Compton wavelength to decrease.
Thus, in our solar system, which sits in a relatively high density region of the 
Universe, the chameleon behavior may cause the $f_R$ field to be undetectable
using our current technologies.

Our code is expected to reproduce this non-linear chameleon behavior that is not 
incorporated into the linear analyses carried out so far.
In order to investigate such effects, we place a spherical top-hat of constant
density at the center of a 200 Mpc simulation box.
The sphere is 20 grid cells in radius, corresponding to 15.6 Mpc, and has
an uniform over-density of $\delta \sim 500$.
Random test masses are placed both inside and outside the sphere, and the 
accelerations of the masses are calculated using the exact solution 
({\gnewton}) and $f_R$ simulations with various field strengths ({\gfr}).
The exact solution of the acceleration increases linearly within the sphere
and decreases as $r^{-2}$ outside the sphere.

In Fig.~\ref{plot:tbf2}, we show the relative difference between {\gfr} and {\gnewton}
as a function of the separation between the center of the spherical blob and the 
test mass.
Outside of the blob, the gravitational force is enhanced as expected from the results
of Fig.~\ref{plot:tbf1}.
However, inside the blob, the $f_R$ field quickly dies down and the gravitational
forces converge to that of $\Lambda$CDM.
Strong $f_R$ fields can penetrate the density field farther, as seen in the case of
$|f_R| = 10^{-3}$ that decays in $\sim 10 h^{-1}$ Mpc.

%-----------------------------------------------------------------------------------------------------
\subsection{Revisiting the Zeldovich pancake}
%-----------------------------------------------------------------------------------------------------
\begin{figure}
  \begin{center}
  \resizebox{84mm}{!}{\includegraphics[angle=0]{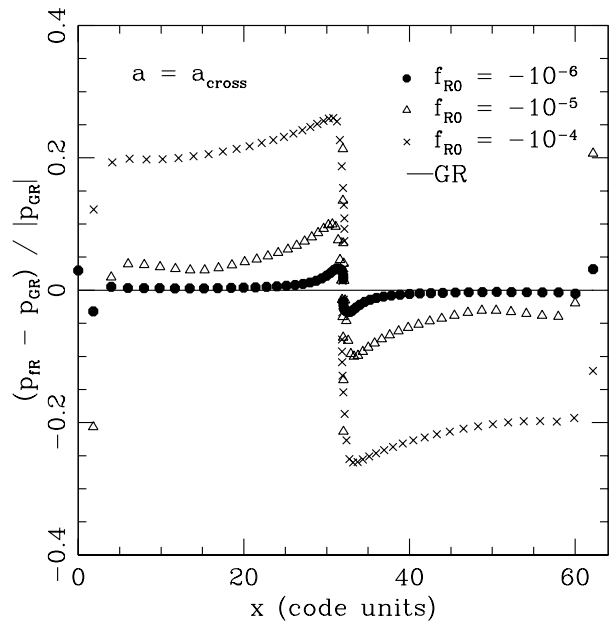}}
  \end{center}
  \caption{The relative differences in particle momenta in Zeldovich pancake
  simulations with varying $f_R$ field strengths.
  The simulation snapshot is taken at $a = a_{\rm cross} = 1$.} 
  \label{plot:zdfr}
\end{figure}
The Zeldovich 1D plane wave collapse offers insights into the differences in the
particle dynamics in the presence of $f(R)$ modifications.
We run three Zeldovich pancake simulations starting with the same cosmology
and initial conditions as in Fig.~\ref{plot:zd1} except for $f(R)$ modifications.
In Fig.~\ref{plot:zdfr}, we show the resulting difference in particle momenta relative
to those of unmodified GR simulation. 
Because of the fifth force due to $f_R$, the sine wave collapses faster as the 
field strength is increased.
In addition, we see the effects of the finite range of the fifth force, which is determined
by the Compton wavelength of the $f_R$ field, Eq.~\ref{eqn:compton}.
In the case of weak field, $f_{R0} = -10^{-4}$, there is no significant enhancement
in the particle momenta at distances greater than $\sim10$ code units ($\sim 16 h^{-1} \rm{Mpc}$)
away from the collapsed pancake, and the range of the fifth force is shown to be
larger for stronger $f_R$ fields.
For the strongest field, the fifth force does not decay sufficiently before the periodic
image of the pancake starts to become important.

The Zeldovich collapses show no sign of the chameleon behavior, even for the
weakest $f_R$ field.
The pancake is essentially a sheet mass and thus does not have an {\it extended} region
of high density.
Therefore, the $f_R$ field does not appreciably decay near the pancake and no 
chameleon behavior is expected.

%-----------------------------------------------------------------------------------------------------
\subsection{Cosmological Power Spectrum}
%-----------------------------------------------------------------------------------------------------
\begin{figure}
  \begin{center}
  \resizebox{84mm}{!}{\includegraphics[angle=0]{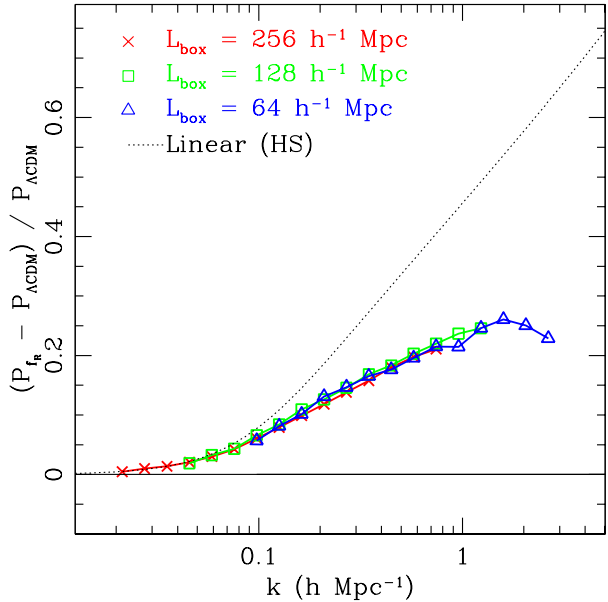}}
  \end{center}
  \caption{Relative differences between the power spectra of $|f_{R0}| = 10^{-5}$
  simulations and unmodified $\Lambda$CDM simulations.
 The linear prediction is taken from Hu \& Sawicki \cite{hu07a}
  and assumes no chameleon behavior.
  The power spectra for each box size are only plotted up to their respective 
  half-Nyquist wave number, which are $0.79 h$ Mpc$^{-1}$, $1.57 h$ Mpc$^{-1}$,
  and $3.14 h$ Mpc$^{-1}$ for L256, L128, and L64 simulations, respectively.
  } 
  \label{plot:psfrac}
\end{figure}

Finally, we apply our code to a realistic cosmological simulation in order to study
the effects of the $f_R$ field to the large scale structures.
We have run $f(R)$ cosmological simulations using the same setup as those found
in \S\ref{subsection:dynamics} and with $f_{R0} = -10^{-5}$.
The strength of the $f_R$ field is chosen to highlight the differences in 
the power spectra.

With our simulation setup having twice as many grid points as particles
per one dimension, we find that at the initial redshift, some of the grid cells
in the simulations have no particles nearby and therefore have zero density.
At early redshifts, the cosmology is very nearly $\Lambda$CDM, and we have 
$R = 8 \pi G \rho$.
Thus, in the empty grid cells, the curvature is almost exactly zero and we have
a formal divergence of the $f_R$ field in such grid cells.
The particle dynamics, however, is governed by the peculiar potential, and therefore
by the curvature and density fields, not directly by the $f_R$ field (see Eq.~\ref{eqn:potcode}).
We have checked that our potential solver and particle trajectory integrators remain 
stable and accurate even when such divergences are present (see Fig.~\ref{plot:dyn1}).
At late times, we also find empty grid cells within large voids.
However, the mean Compton wavelength of the $f_R$ field becomes 
sufficiently large and smoothes out the $f_R$ field, thereby avoiding any 
potential divergences in the field at late times.

%As expected from the results of \S \ref{subsection:tbf}, the power spectra
%are enhanced compared to the respective $\Lambda$CDM power spectra.
Fig.~\ref{plot:psfrac} shows the relative differences between the power spectra
with and without $f(R)$ modifications, along with the linear prediction for
the power spectrum enhancement taken from \cite{hu07a}.
The linear prediction assumes no chameleon behavior and thus is expected
to overestimate the power spectrum enhancement when there is strong
suppression of $f_R$ field in high density regions.
The simulated power spectra match the linear predictions at large scales,
where non-linear effects, such as the chameleon mechanism, is expected
to be negligible.  
The smallest simulation box, $L_{\rm box} = 64 h^{-1}$ Mpc, does not have 
sufficient statistics to accurately represent the largest modes.
At scales smaller than $k \sim 0.1 h$ Mpc, we see a striking deviation of the
simulations from the linear prediction, showing clear signs the previously
unquantified effects of non-linear $f(R)$ physics.

More detailed comparison of the power spectra are carried out in the companion
paper.

%-----------------------------------------------------------------------------------------------------
%\subsection{Initial Conditions}
%
%\textcolor{red}{I'm debating whether this section should be included}
%-----------------------------------------------------------------------------------------------------

%%%%%%%%%%%%%%%%%%%%%%%%%%%%%%%%%%%%%%%%%
\section{conclusion} \label{sec:conclusion}
%%%%%%%%%%%%%%%%%%%%%%%%%%%%%%%%%%%%%%%%%
In this paper, we have introduced a method to construct a cosmological
N-body simulation with a class of viable $f(R)$ modifications to general
relativity. 
Our simulation code, implemented in standard C++, solves the
HS model of $f(R)$ modification using particle mesh method, particularly
suited because the scalar degree of freedom in the model is well 
represented as a scalar field (i.e., a mesh) coupled to the Newtonian
potential.
The non-linear field equation for $f_R$ is solved on the mesh using an efficient
multigrid scheme that is shown to converge robustly to the desired solution.
The code is also shown to reproduce the chameleon behavior of the HS
model.

The code should allow us to gain insight into the quasi-nonliear physics of
$f(R)$ modified gravity models.  
In general, gravitational forces are enhanced under the HS model, and we
show that the resulting power spectra are also enhanced compared to 
the $\Lambda$CDM power spectrum.
We further show hints that the power spectrum enhancement is smaller
than the linear predictions due to the chameleon mechanism that has not
so far been modelled.
In a companion paper \citep{oyaizu08b}, we further study the effects of the HS 
model on the dark matter power spectrum and its cosmological implications.
In addition, effects on the cluster mass function and weak lensing statistics
are potentially interesting and should be investigated in the future.

Lastly, the work presented here can be used as a general framework
for non-linear simulation of any minimally coupled scalar field model
of gravity, provided that the resulting field equations are sufficiently
stable under relaxation and multigrid schemes.
In particular, an interesting exercise might be to apply this framework
to the Dvali, Gabadadze, and Porrati (DGP) model \citep{dvali00a}.

\section{Acknowledgments}

\ackntext{We would like to thank Wayne Hu, Marcos Lima, Justin Khoury, 
Josh Frieman, and Jeremy Tinker for insightful and useful
discussions, and Andrey Kravtsov for guiding the authors during the 
implementation and testing of the simulation code.
This work was supported in part by the Kavli Institute for Cosmological
 Physics (KICP) at the University of Chicago through grants NSF PHY-0114422 and
 NSF PHY-0551142 and an endowment from the Kavli Foundation and its 
founder Fred Kavli.
HO was additionally supported by the NSF grants AST-0239759, AST-0507666, 
and AST-0708154 at the University of Chicago.
Some of the computations used in this work have been performed on the Joint
Fermilab - KICP Supercomputing Cluster, supported by grants from Fermilab,
Kavli Insititute for Cosmological Physics, and the University of Chicago.
This research has made use of NASA's Astrophysics Data System and arXiv.org
preprint service funded by Cornell University and the NSF.
}

\appendix
\section{Details of the discretization} \label{appendix:discretization}
In our implementation, Eq.~(\ref{eqn:diffu}) is discretized much like a standard
variable coefficient Poisson equation.  
We let $a = e^u$, $a_{i+\frac{1}{2},j,k} = (a_{j,j,k} + a_{j+1,j,k}) / 2$, and $a_{i-\frac{1}{2},j,k} = (a_{i,j,k} + a_{i-1,j,k})/2$.
Then, we have
\begin{widetext}
\begin{eqnarray}
\lsc &=& \frac{1}{h^2} \left(a_{i-\frac{1}{2},j,k} u_{i-1,j,k} - (a_{i-\frac{1}{2},j,k}+a_{i+\frac{1}{2},j,k}) u_{i,j,k} + a_{i+\frac{1}{2},j,k} u_{i+1,j,k}  \right) \nonumber \\
&&+\frac{1}{h^2} \left(a_{i,j-\frac{1}{2},k} u_{i,j-1,k} - (a_{i,j-\frac{1}{2},k}+a_{i,j+\frac{1}{2},k}) u_{i,j,k} + a_{i,j+\frac{1}{2},k} u_{i,j+1,k}  \right) \nonumber \\
&&+\frac{1}{h^2} \left(a_{i,j,k-\frac{1}{2}} u_{i,j,k-1} - (a_{i,j,k-\frac{1}{2}}+a_{i,j,k+\frac{1}{2}}) u_{i,j,k} + a_{i,j,k+\frac{1}{2}} u_{i,j,k+1}  \right)  \nonumber \\
&&- \frac{\Omega_{M,0}}{a \tilde{c}^2 \bar{f_R}}\left[  \frac{ R(\bar{f_R}e^u_{i,j,k}) - \bar{R}}{3} -  \delta_{i,j,k} \right] . \label{eqn:ufrdiff}
\end{eqnarray}
\end{widetext}
We find that the above discretization performs better than discretizing the equation 
after expanding out the derivative operators.
Even using this discretization, we find that the iterations can diverge
if the initial guess is sufficiently poor.
We find that using variable, and often smaller, Newton-Raphson step sizes 
(i.e., successive \textit{under}-relaxation) can
effectively avoid such divergences until the trial solution is sufficiently
close to the true solution.

\section{FAS implementation} \label{appendix:multigrid}
In our FAS implementation, we use the full-weighting restriction operator ($I^{2h}_h$)
and the bilinear interpolation operator ($I^h_{2h}$) for prolongation.
The full-weighting operator, in 3D, is given by
\begin{eqnarray}
u^{2h}_{i,j,k} = \sum_{\alpha,\beta,\gamma=-1}^{1} \sigma_{\alpha,\beta,\gamma} u^h_{2i+\alpha,2j+\beta,2k+\gamma}
\end{eqnarray}
where
\begin{eqnarray}
\sigma_{\alpha,\beta,\gamma} &=& \frac{1}{8} \ \ \text{if} \ \  |\alpha|+|\beta|+|\gamma| = 0 \nonumber \\
\sigma_{\alpha,\beta,\gamma} &=& \frac{1}{16} \ \ \text{if} \ \  |\alpha|+|\beta|+|\gamma| = 1 \nonumber \\
\sigma_{\alpha,\beta,\gamma} &=& \frac{1}{32} \ \ \text{if} \ \  |\alpha|+|\beta|+|\gamma| = 2 \nonumber \\
\sigma_{\alpha,\beta,\gamma} &=& \frac{1}{64} \ \ \text{if} \ \  |\alpha|+|\beta|+|\gamma| = 3 \nonumber.
\end{eqnarray}
Note that the $2h$ superscript denotes a variable defined on the coarse grid 
(i.e., a grid with spacing of $2h$) and the $h$ superscript denotes a variable
defined on the fine grid.
The bilinear interpolation operator is given by
\begin{widetext}
\begin{eqnarray}
u^{h}_{2i,2j,2k} &=& u^{2h}_{i,j,k} \nonumber \\
u^{h}_{2i+1,2j,2k} &=& \frac{1}{2}(u^{2h}_{i,j,k} + u^{2h}_{i+1,j,k}) \nonumber \\
u^{h}_{2i,2j+1,2k} &=& \frac{1}{2}(u^{2h}_{i,j,k} + u^{2h}_{i,j+1,k}) \nonumber \\
u^{h}_{2i,2j,2k+1} &=& \frac{1}{2}(u^{2h}_{i,j,k} + u^{2h}_{i,j,k+1}) \nonumber \\
u^{h}_{2i+1,2j+1,2k} &=& \frac{1}{4}(u^{2h}_{i,j,k} + u^{2h}_{i+1,j,k} + u^{2h}_{i,j+1,k} + u^{2h}_{i+1,j+1,k}) \nonumber \\
u^{h}_{2i,2j+1,2k+1} &=& \frac{1}{4}(u^{2h}_{i,j,k} + u^{2h}_{i,j+1,k} + u^{2h}_{i,j,k+1} + u^{2h}_{i,j+1,k+1}) \nonumber \\
u^{h}_{2i+1,2j,2k+1} &=& \frac{1}{4}(u^{2h}_{i,j,k} + u^{2h}_{i+1,j,k} + u^{2h}_{i,j,k+1} + u^{2h}_{i+1,j,k+1}) \nonumber \\
u^{h}_{2i+1,2j+1,2k+1} &=& \frac{1}{8}(u^{2h}_{i,j,k} + u^{2h}_{i+1,j,k} + u^{2h}_{i,j+1,k} + u^{2h}_{i,j,k+1} \\
&& +u^{2h}_{i+1,j+1,k} + u^{2h}_{i+1,j,k+1} + u^{2h}_{i,j+1,k+1} + u^{2h}_{i+1,j+1,k+1}). \nonumber 
\end{eqnarray}
\end{widetext}
Other choices of restriction and prolongation operators are certainly possible.
However, we find that the above choices result in the fastest and most stable
FAS implementation for this particular problem.
%%%%%%%%%%%%%%%%%%%%%%%%%%%%%%%%%%%%%%%%%
\bibliography{methodpaper}
%%%%%%%%%%%%%%%%%%%%%%%%%%%%%%%%%%%%%%%%%

\end{document}